\documentstyle[12pt,psfig]{article}

\begin{document}

\setcounter{page}{1}

\rightline{ISU--NP--94--14}
\vskip .5in
\centerline{\large {\bf Massive Schwinger Model for Small Fermion Mass}
\footnote{Based on a talk presented by J.P. Vary at ``Theory of Hadrons and
Light-Front QCD'', Polona Zgorzelisko, Poland, August 1994.}}

\vskip.3in
\centerline{J.P.~Vary and T.J.~Fields}
\centerline{{\it Physics Department, Iowa State University of
Science and Technology}}
\centerline{{\it Ames, IA  50011}}
\vskip.2 in
\centerline{H.J.~Pirner}
\centerline{{\it Inst. f. Theoret. Physik, Univ. Heidelberg,}}
\centerline{{\it Philosophenweg 19, D-69120 Heidelberg, Germany}}

\vskip.6 in
\centerline{\bf Abstract}
\vskip.1in
We examine QED$_{1+1}$ (the massive Schwinger model) to extend existing
perturbation results in $m_f/m_b$, and compare our results with lattice
and DLCQ calculations.
\vskip.1in

\begin{figure}
\centerline{\psfig{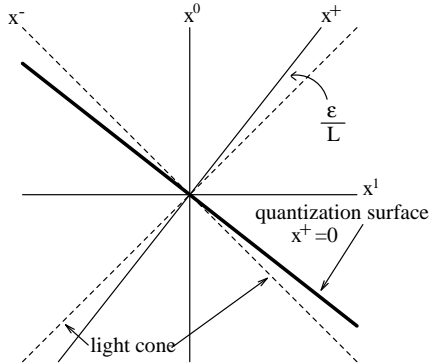}}
\caption{A diagram of our coordinate system}
\label{coords}
\end{figure}

\vskip.3in
{\bf 1. Introduction}
\vskip.1in
We start with the standard Lagrangian of QED, and work in a coordinate
system that is `near' the light front$^{1,2}$:
\begin{eqnarray}
x^{+} & = &\frac{1}{\sqrt{2}}\left [\left ( 1+\frac{\epsilon}{L}\right ) x^{0}
+\left ( 1-\frac{\epsilon}{L}\right ) x^{1}\right ]\\
x^{-} & = & \frac{1}{\sqrt{2}}(x^{0}-x^{1})
\label{coord}
\end{eqnarray}
We interpret $x^+$ as our time variable, and $x^-$ as our
spatial variable.
We will discretize the problem by
putting it in a box $x^{-} = [0,L]$ and require periodic
boundary conditions.
Note that with this choice of coordinate system, the two ends
of our box
are separated by a spacelike separation $ds^{2}=-2\epsilon L$, and
no conflict with causality arises by imposing our boundary conditions
. In other words, it allows us to specify our initial
conditions on the spacelike
surface $x^{+}=0$ (fig. \ref{coords}), while allowing the recovery
of the `usual' light
front variables by letting $\epsilon \to 0$ for a fixed $L$.

In `pure' light front coordinates, the negative momentum states
are eliminated because, as the dispersion relation
\begin{equation}
p_{+}=\frac{m^2}{2p_{-}}
\end{equation}
shows, a negative momentum state is a state of negative energy.
In our coordinates, we can no longer neglect the negative momentum
states.  The dispersion relation now reads,
\begin{equation}
p_{+}=\frac{L}{2\epsilon}\left (- p_{-} \pm \sqrt{(p_{-})^2+
\frac{2m^2 \epsilon}{L}}\right )
\end{equation}
showing that negative momentum modes can contribute to
positive energy states.

As the axial gauge $A_{-}=0$ is inconsistent with boundary
conditions on
this finite interval, we pick the light front Couloumb gauge
$\partial_{-}A_{-}=0$, which poses no such problems. The need to
restrict
the problem to the charge zero sector is well known, and we choose
a heat-kernal regularization for the problem, which is designed to be
gauge invariant.

We can completely diagonalize the Hamiltonian in an appropriately
defined set of bosonic operators.  After manipulation to this form, we
can set $m_f=0$, and in the continuum limit, we
obtain standard Schwinger model results:
\begin{itemize}
\item{Physical states are non--interacting Schwinger bosons}
\item{Mass of Schwinger boson is $m_b^{(0)} =g/ \sqrt{\pi}$.}
\item{$\theta$ vacuum with condensate $\langle \theta |\bar{\psi}\psi| \theta
\rangle = \frac{1}{2 \pi}m_b^{(0)}
e^{c}\cos{\theta}$}
\end{itemize}
where $c$ is Euler's constant ($c=0.5772\ldots$).
\vskip.3in
\newpage
{\bf 2. Mass Perturbation}
\vskip.1in

The idea is to simply let the quark have a small mass $m_f$, and
treat the mass term in the Hamiltonian as a perturbation.
Therefore the perturbation will be in dimensionless units of
$(m_f/g)$ and hence the region of small mass directly
corresponds to the region of strong coupling.

Using our bosonized Hamiltonian, we recover known first order results
for the change in the mass of the Schwinger boson$^1$:
\begin{equation}
\left( \delta m_{b}^{2} \right )^{(1)}
= -4\pi m_{f}\langle \theta |\bar{\psi}\psi| \theta \rangle =
-2m_{f}m_{b}^{(0)}e^{c}\cos{\theta}
\end{equation}

\begin{figure}
\centerline{\psfig{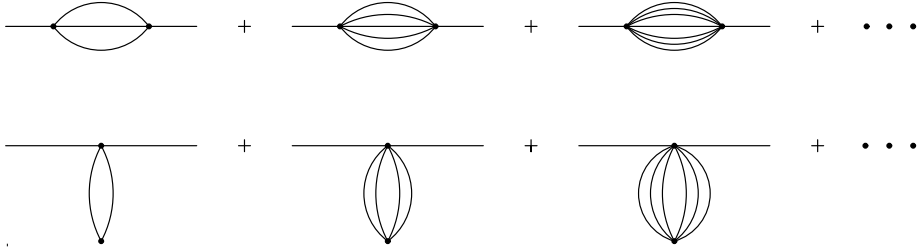}}
\caption{A diagrammatic representation of the expansion
for the shift in the mass of
the Schwinger boson to second order when quantized in the near light-front
coordinates.}
\label{expansion}
\end{figure}
In second order, we obtain new results.
We can write our second order series for $\delta m_b^2$
in diagrammatic form (fig. \ref{expansion}).
There are a few important things to note in this expansion
\begin{itemize}
\item{Each term in the expansion is finite.}
\item{Each row, summed separately, diverges.}
\item{Divergences {\it cancel} when {\it all} diagrams are summed.}
\item{High Fock states are very important: all the diagrams shown
  explicitly in figure \ref{expansion} contribute only $40\%$
  to the total sum.}
\end{itemize}
The non-vanishing contributions of the diagrams from
the lower row of the figure is a direct
consequence of the presence of negative momentum states in
our choice of coordinate system$^{3,4}$

Through this evaluation, we identify an expansion parameter:
\begin{equation}
\beta = 2e^{c}\frac{m_f}{m_b^{(0)}} \propto m_f \langle
\theta |\bar{\psi}\psi| \theta \rangle
\end{equation}
with which we can write the results through second order as follows:
\begin{equation}
m_{b}^2=\left( m_b^{(0)}\right )^2\left [1 + \beta +
\frac{A}{2}\beta^{2}\right ]^{2}
\end{equation}
with A defined by:
\begin{equation}
A=-\int_{0}^{\infty}\rho \, d\rho
\left [ \left \{ \sinh{2K_0(\rho)}-2K_0(\rho)\right \} I_{0}
(\rho)+1-\cosh{2K_0(\rho)} \right ].
\end{equation}
As potentially nasty as this integrand looks (The modified Bessel functions
$K_0(\rho)$ and $I_0(\rho)$ diverge as
$\rho \to 0$, and as $\rho \to \infty$, respectively), it
is actually well behaved, and $A$ can be evaluated
numerically to be $0.5339\ldots$.

\begin{figure}
%\centerline{\psfig{figure=comparison.ps}}
\centering
% GNUPLOT: LaTeX picture
\setlength{\unitlength}{0.240900pt}
\ifx\plotpoint\undefined\newsavebox{\plotpoint}\fi
\sbox{\plotpoint}{\rule[-0.200pt]{0.400pt}{0.400pt}}%
\begin{picture}(1500,900)(0,0)
\font\gnuplot=cmr10 at 10pt
\gnuplot
\sbox{\plotpoint}{\rule[-0.200pt]{0.400pt}{0.400pt}}%
\put(220.0,113.0){\rule[-0.200pt]{4.818pt}{0.400pt}}
\put(198,113){\makebox(0,0)[r]{1}}
\put(1416.0,113.0){\rule[-0.200pt]{4.818pt}{0.400pt}}
\put(220.0,215.0){\rule[-0.200pt]{4.818pt}{0.400pt}}
\put(198,215){\makebox(0,0)[r]{1.2}}
\put(1416.0,215.0){\rule[-0.200pt]{4.818pt}{0.400pt}}
\put(220.0,317.0){\rule[-0.200pt]{4.818pt}{0.400pt}}
\put(198,317){\makebox(0,0)[r]{1.4}}
\put(1416.0,317.0){\rule[-0.200pt]{4.818pt}{0.400pt}}
\put(220.0,419.0){\rule[-0.200pt]{4.818pt}{0.400pt}}
\put(198,419){\makebox(0,0)[r]{1.6}}
\put(1416.0,419.0){\rule[-0.200pt]{4.818pt}{0.400pt}}
\put(220.0,520.0){\rule[-0.200pt]{4.818pt}{0.400pt}}
\put(198,520){\makebox(0,0)[r]{1.8}}
\put(1416.0,520.0){\rule[-0.200pt]{4.818pt}{0.400pt}}
\put(220.0,622.0){\rule[-0.200pt]{4.818pt}{0.400pt}}
\put(198,622){\makebox(0,0)[r]{2}}
\put(1416.0,622.0){\rule[-0.200pt]{4.818pt}{0.400pt}}
\put(220.0,724.0){\rule[-0.200pt]{4.818pt}{0.400pt}}
\put(198,724){\makebox(0,0)[r]{2.2}}
\put(1416.0,724.0){\rule[-0.200pt]{4.818pt}{0.400pt}}
\put(220.0,826.0){\rule[-0.200pt]{4.818pt}{0.400pt}}
\put(198,826){\makebox(0,0)[r]{2.4}}
\put(1416.0,826.0){\rule[-0.200pt]{4.818pt}{0.400pt}}
\put(220.0,113.0){\rule[-0.200pt]{0.400pt}{4.818pt}}
\put(220,68){\makebox(0,0){$2^{-8}$}}
\put(220.0,857.0){\rule[-0.200pt]{0.400pt}{4.818pt}}
\put(435.0,113.0){\rule[-0.200pt]{0.400pt}{4.818pt}}
\put(435,68){\makebox(0,0){$2^{-6}$}}
\put(435.0,857.0){\rule[-0.200pt]{0.400pt}{4.818pt}}
\put(650.0,113.0){\rule[-0.200pt]{0.400pt}{4.818pt}}
\put(650,68){\makebox(0,0){$2^{-4}$}}
\put(650.0,857.0){\rule[-0.200pt]{0.400pt}{4.818pt}}
\put(864.0,113.0){\rule[-0.200pt]{0.400pt}{4.818pt}}
\put(864,68){\makebox(0,0){$2^{-2}$}}
\put(864.0,857.0){\rule[-0.200pt]{0.400pt}{4.818pt}}
\put(1079.0,113.0){\rule[-0.200pt]{0.400pt}{4.818pt}}
\put(1079,68){\makebox(0,0){$2^{0}$}}
\put(1079.0,857.0){\rule[-0.200pt]{0.400pt}{4.818pt}}
\put(1294.0,113.0){\rule[-0.200pt]{0.400pt}{4.818pt}}
\put(1294,68){\makebox(0,0){$2^{2}$}}
\put(1294.0,857.0){\rule[-0.200pt]{0.400pt}{4.818pt}}
\put(220.0,113.0){\rule[-0.200pt]{292.934pt}{0.400pt}}
\put(1436.0,113.0){\rule[-0.200pt]{0.400pt}{184.048pt}}
\put(220.0,877.0){\rule[-0.200pt]{292.934pt}{0.400pt}}
\put(15,495){\makebox(0,0){$\frac{m_b}{\sqrt{m_f^2+g^2 / \pi}}$}}
\put(828,3){\makebox(0,0){$m_f/g$}}
\put(1187,317){\makebox(0,0)[l]{first order}}
\put(667,622){\makebox(0,0)[l]{second order}}
\put(406,801){\makebox(0,0)[l]{lattice}}
\put(406,750){\makebox(0,0)[l]{DLCQ}}
\put(220.0,113.0){\rule[-0.200pt]{0.400pt}{184.048pt}}
\multiput(1246.47,289.92)(-0.638,-0.499){149}{\rule{0.611pt}{0.120pt}}
\multiput(1247.73,290.17)(-95.733,-76.000){2}{\rule{0.305pt}{0.400pt}}
\put(1152,215){\vector(-4,-3){0}}
\multiput(795.00,595.92)(0.636,-0.499){151}{\rule{0.609pt}{0.120pt}}
\multiput(795.00,596.17)(96.736,-77.000){2}{\rule{0.305pt}{0.400pt}}
\put(893,520){\vector(4,-3){0}}
\put(220,119){\usebox{\plotpoint}}
\put(220,118.67){\rule{2.891pt}{0.400pt}}
\multiput(220.00,118.17)(6.000,1.000){2}{\rule{1.445pt}{0.400pt}}
\put(245,119.67){\rule{2.891pt}{0.400pt}}
\multiput(245.00,119.17)(6.000,1.000){2}{\rule{1.445pt}{0.400pt}}
\put(257,120.67){\rule{2.891pt}{0.400pt}}
\multiput(257.00,120.17)(6.000,1.000){2}{\rule{1.445pt}{0.400pt}}
\put(232.0,120.0){\rule[-0.200pt]{3.132pt}{0.400pt}}
\put(281,121.67){\rule{3.132pt}{0.400pt}}
\multiput(281.00,121.17)(6.500,1.000){2}{\rule{1.566pt}{0.400pt}}
\put(294,122.67){\rule{2.891pt}{0.400pt}}
\multiput(294.00,122.17)(6.000,1.000){2}{\rule{1.445pt}{0.400pt}}
\put(306,123.67){\rule{2.891pt}{0.400pt}}
\multiput(306.00,123.17)(6.000,1.000){2}{\rule{1.445pt}{0.400pt}}
\put(318,124.67){\rule{3.132pt}{0.400pt}}
\multiput(318.00,124.17)(6.500,1.000){2}{\rule{1.566pt}{0.400pt}}
\put(331,125.67){\rule{2.891pt}{0.400pt}}
\multiput(331.00,125.17)(6.000,1.000){2}{\rule{1.445pt}{0.400pt}}
\put(343,126.67){\rule{2.891pt}{0.400pt}}
\multiput(343.00,126.17)(6.000,1.000){2}{\rule{1.445pt}{0.400pt}}
\put(355,127.67){\rule{2.891pt}{0.400pt}}
\multiput(355.00,127.17)(6.000,1.000){2}{\rule{1.445pt}{0.400pt}}
\put(367,129.17){\rule{2.700pt}{0.400pt}}
\multiput(367.00,128.17)(7.396,2.000){2}{\rule{1.350pt}{0.400pt}}
\put(380,130.67){\rule{2.891pt}{0.400pt}}
\multiput(380.00,130.17)(6.000,1.000){2}{\rule{1.445pt}{0.400pt}}
\put(392,132.17){\rule{2.500pt}{0.400pt}}
\multiput(392.00,131.17)(6.811,2.000){2}{\rule{1.250pt}{0.400pt}}
\put(404,133.67){\rule{3.132pt}{0.400pt}}
\multiput(404.00,133.17)(6.500,1.000){2}{\rule{1.566pt}{0.400pt}}
\put(417,135.17){\rule{2.500pt}{0.400pt}}
\multiput(417.00,134.17)(6.811,2.000){2}{\rule{1.250pt}{0.400pt}}
\put(429,137.17){\rule{2.500pt}{0.400pt}}
\multiput(429.00,136.17)(6.811,2.000){2}{\rule{1.250pt}{0.400pt}}
\put(441,139.17){\rule{2.500pt}{0.400pt}}
\multiput(441.00,138.17)(6.811,2.000){2}{\rule{1.250pt}{0.400pt}}
\put(453,141.17){\rule{2.700pt}{0.400pt}}
\multiput(453.00,140.17)(7.396,2.000){2}{\rule{1.350pt}{0.400pt}}
\multiput(466.00,143.61)(2.472,0.447){3}{\rule{1.700pt}{0.108pt}}
\multiput(466.00,142.17)(8.472,3.000){2}{\rule{0.850pt}{0.400pt}}
\multiput(478.00,146.61)(2.472,0.447){3}{\rule{1.700pt}{0.108pt}}
\multiput(478.00,145.17)(8.472,3.000){2}{\rule{0.850pt}{0.400pt}}
\put(490,149.17){\rule{2.700pt}{0.400pt}}
\multiput(490.00,148.17)(7.396,2.000){2}{\rule{1.350pt}{0.400pt}}
\multiput(503.00,151.60)(1.651,0.468){5}{\rule{1.300pt}{0.113pt}}
\multiput(503.00,150.17)(9.302,4.000){2}{\rule{0.650pt}{0.400pt}}
\multiput(515.00,155.61)(2.472,0.447){3}{\rule{1.700pt}{0.108pt}}
\multiput(515.00,154.17)(8.472,3.000){2}{\rule{0.850pt}{0.400pt}}
\multiput(527.00,158.60)(1.651,0.468){5}{\rule{1.300pt}{0.113pt}}
\multiput(527.00,157.17)(9.302,4.000){2}{\rule{0.650pt}{0.400pt}}
\multiput(539.00,162.60)(1.797,0.468){5}{\rule{1.400pt}{0.113pt}}
\multiput(539.00,161.17)(10.094,4.000){2}{\rule{0.700pt}{0.400pt}}
\multiput(552.00,166.60)(1.651,0.468){5}{\rule{1.300pt}{0.113pt}}
\multiput(552.00,165.17)(9.302,4.000){2}{\rule{0.650pt}{0.400pt}}
\multiput(564.00,170.60)(1.651,0.468){5}{\rule{1.300pt}{0.113pt}}
\multiput(564.00,169.17)(9.302,4.000){2}{\rule{0.650pt}{0.400pt}}
\multiput(576.00,174.59)(1.267,0.477){7}{\rule{1.060pt}{0.115pt}}
\multiput(576.00,173.17)(9.800,5.000){2}{\rule{0.530pt}{0.400pt}}
\multiput(588.00,179.59)(1.123,0.482){9}{\rule{0.967pt}{0.116pt}}
\multiput(588.00,178.17)(10.994,6.000){2}{\rule{0.483pt}{0.400pt}}
\multiput(601.00,185.59)(1.267,0.477){7}{\rule{1.060pt}{0.115pt}}
\multiput(601.00,184.17)(9.800,5.000){2}{\rule{0.530pt}{0.400pt}}
\multiput(613.00,190.59)(0.874,0.485){11}{\rule{0.786pt}{0.117pt}}
\multiput(613.00,189.17)(10.369,7.000){2}{\rule{0.393pt}{0.400pt}}
\multiput(625.00,197.59)(1.123,0.482){9}{\rule{0.967pt}{0.116pt}}
\multiput(625.00,196.17)(10.994,6.000){2}{\rule{0.483pt}{0.400pt}}
\multiput(638.00,203.59)(0.758,0.488){13}{\rule{0.700pt}{0.117pt}}
\multiput(638.00,202.17)(10.547,8.000){2}{\rule{0.350pt}{0.400pt}}
\multiput(650.00,211.59)(0.874,0.485){11}{\rule{0.786pt}{0.117pt}}
\multiput(650.00,210.17)(10.369,7.000){2}{\rule{0.393pt}{0.400pt}}
\multiput(662.00,218.59)(0.758,0.488){13}{\rule{0.700pt}{0.117pt}}
\multiput(662.00,217.17)(10.547,8.000){2}{\rule{0.350pt}{0.400pt}}
\multiput(674.00,226.59)(0.728,0.489){15}{\rule{0.678pt}{0.118pt}}
\multiput(674.00,225.17)(11.593,9.000){2}{\rule{0.339pt}{0.400pt}}
\multiput(687.00,235.58)(0.600,0.491){17}{\rule{0.580pt}{0.118pt}}
\multiput(687.00,234.17)(10.796,10.000){2}{\rule{0.290pt}{0.400pt}}
\multiput(699.00,245.58)(0.600,0.491){17}{\rule{0.580pt}{0.118pt}}
\multiput(699.00,244.17)(10.796,10.000){2}{\rule{0.290pt}{0.400pt}}
\multiput(711.00,255.58)(0.590,0.492){19}{\rule{0.573pt}{0.118pt}}
\multiput(711.00,254.17)(11.811,11.000){2}{\rule{0.286pt}{0.400pt}}
\multiput(724.00,266.58)(0.543,0.492){19}{\rule{0.536pt}{0.118pt}}
\multiput(724.00,265.17)(10.887,11.000){2}{\rule{0.268pt}{0.400pt}}
\multiput(736.58,277.00)(0.492,0.539){21}{\rule{0.119pt}{0.533pt}}
\multiput(735.17,277.00)(12.000,11.893){2}{\rule{0.400pt}{0.267pt}}
\multiput(748.58,290.00)(0.492,0.539){21}{\rule{0.119pt}{0.533pt}}
\multiput(747.17,290.00)(12.000,11.893){2}{\rule{0.400pt}{0.267pt}}
\multiput(760.58,303.00)(0.493,0.536){23}{\rule{0.119pt}{0.531pt}}
\multiput(759.17,303.00)(13.000,12.898){2}{\rule{0.400pt}{0.265pt}}
\multiput(773.58,317.00)(0.492,0.582){21}{\rule{0.119pt}{0.567pt}}
\multiput(772.17,317.00)(12.000,12.824){2}{\rule{0.400pt}{0.283pt}}
\multiput(785.58,331.00)(0.492,0.669){21}{\rule{0.119pt}{0.633pt}}
\multiput(784.17,331.00)(12.000,14.685){2}{\rule{0.400pt}{0.317pt}}
\multiput(797.58,347.00)(0.493,0.616){23}{\rule{0.119pt}{0.592pt}}
\multiput(796.17,347.00)(13.000,14.771){2}{\rule{0.400pt}{0.296pt}}
\multiput(810.58,363.00)(0.492,0.669){21}{\rule{0.119pt}{0.633pt}}
\multiput(809.17,363.00)(12.000,14.685){2}{\rule{0.400pt}{0.317pt}}
\multiput(822.58,379.00)(0.492,0.755){21}{\rule{0.119pt}{0.700pt}}
\multiput(821.17,379.00)(12.000,16.547){2}{\rule{0.400pt}{0.350pt}}
\multiput(834.58,397.00)(0.492,0.755){21}{\rule{0.119pt}{0.700pt}}
\multiput(833.17,397.00)(12.000,16.547){2}{\rule{0.400pt}{0.350pt}}
\multiput(846.58,415.00)(0.493,0.734){23}{\rule{0.119pt}{0.685pt}}
\multiput(845.17,415.00)(13.000,17.579){2}{\rule{0.400pt}{0.342pt}}
\multiput(859.58,434.00)(0.492,0.755){21}{\rule{0.119pt}{0.700pt}}
\multiput(858.17,434.00)(12.000,16.547){2}{\rule{0.400pt}{0.350pt}}
\multiput(871.58,452.00)(0.492,0.841){21}{\rule{0.119pt}{0.767pt}}
\multiput(870.17,452.00)(12.000,18.409){2}{\rule{0.400pt}{0.383pt}}
\multiput(883.58,472.00)(0.493,0.734){23}{\rule{0.119pt}{0.685pt}}
\multiput(882.17,472.00)(13.000,17.579){2}{\rule{0.400pt}{0.342pt}}
\multiput(896.58,491.00)(0.492,0.798){21}{\rule{0.119pt}{0.733pt}}
\multiput(895.17,491.00)(12.000,17.478){2}{\rule{0.400pt}{0.367pt}}
\multiput(908.58,510.00)(0.492,0.798){21}{\rule{0.119pt}{0.733pt}}
\multiput(907.17,510.00)(12.000,17.478){2}{\rule{0.400pt}{0.367pt}}
\multiput(920.58,529.00)(0.492,0.755){21}{\rule{0.119pt}{0.700pt}}
\multiput(919.17,529.00)(12.000,16.547){2}{\rule{0.400pt}{0.350pt}}
\multiput(932.58,547.00)(0.493,0.655){23}{\rule{0.119pt}{0.623pt}}
\multiput(931.17,547.00)(13.000,15.707){2}{\rule{0.400pt}{0.312pt}}
\multiput(945.58,564.00)(0.492,0.712){21}{\rule{0.119pt}{0.667pt}}
\multiput(944.17,564.00)(12.000,15.616){2}{\rule{0.400pt}{0.333pt}}
\multiput(957.58,581.00)(0.492,0.625){21}{\rule{0.119pt}{0.600pt}}
\multiput(956.17,581.00)(12.000,13.755){2}{\rule{0.400pt}{0.300pt}}
\multiput(969.58,596.00)(0.493,0.536){23}{\rule{0.119pt}{0.531pt}}
\multiput(968.17,596.00)(13.000,12.898){2}{\rule{0.400pt}{0.265pt}}
\multiput(982.00,610.58)(0.496,0.492){21}{\rule{0.500pt}{0.119pt}}
\multiput(982.00,609.17)(10.962,12.000){2}{\rule{0.250pt}{0.400pt}}
\multiput(994.00,622.58)(0.543,0.492){19}{\rule{0.536pt}{0.118pt}}
\multiput(994.00,621.17)(10.887,11.000){2}{\rule{0.268pt}{0.400pt}}
\multiput(1006.00,633.59)(0.669,0.489){15}{\rule{0.633pt}{0.118pt}}
\multiput(1006.00,632.17)(10.685,9.000){2}{\rule{0.317pt}{0.400pt}}
\multiput(1018.00,642.59)(0.824,0.488){13}{\rule{0.750pt}{0.117pt}}
\multiput(1018.00,641.17)(11.443,8.000){2}{\rule{0.375pt}{0.400pt}}
\multiput(1031.00,650.59)(1.033,0.482){9}{\rule{0.900pt}{0.116pt}}
\multiput(1031.00,649.17)(10.132,6.000){2}{\rule{0.450pt}{0.400pt}}
\multiput(1043.00,656.60)(1.651,0.468){5}{\rule{1.300pt}{0.113pt}}
\multiput(1043.00,655.17)(9.302,4.000){2}{\rule{0.650pt}{0.400pt}}
\multiput(1055.00,660.61)(2.695,0.447){3}{\rule{1.833pt}{0.108pt}}
\multiput(1055.00,659.17)(9.195,3.000){2}{\rule{0.917pt}{0.400pt}}
\put(1068,662.67){\rule{2.891pt}{0.400pt}}
\multiput(1068.00,662.17)(6.000,1.000){2}{\rule{1.445pt}{0.400pt}}
\put(269.0,122.0){\rule[-0.200pt]{2.891pt}{0.400pt}}
\put(1092,662.67){\rule{2.891pt}{0.400pt}}
\multiput(1092.00,663.17)(6.000,-1.000){2}{\rule{1.445pt}{0.400pt}}
\put(1104,661.67){\rule{3.132pt}{0.400pt}}
\multiput(1104.00,662.17)(6.500,-1.000){2}{\rule{1.566pt}{0.400pt}}
\multiput(1117.00,660.95)(2.472,-0.447){3}{\rule{1.700pt}{0.108pt}}
\multiput(1117.00,661.17)(8.472,-3.000){2}{\rule{0.850pt}{0.400pt}}
\multiput(1129.00,657.95)(2.472,-0.447){3}{\rule{1.700pt}{0.108pt}}
\multiput(1129.00,658.17)(8.472,-3.000){2}{\rule{0.850pt}{0.400pt}}
\multiput(1141.00,654.94)(1.651,-0.468){5}{\rule{1.300pt}{0.113pt}}
\multiput(1141.00,655.17)(9.302,-4.000){2}{\rule{0.650pt}{0.400pt}}
\multiput(1153.00,650.94)(1.797,-0.468){5}{\rule{1.400pt}{0.113pt}}
\multiput(1153.00,651.17)(10.094,-4.000){2}{\rule{0.700pt}{0.400pt}}
\multiput(1166.00,646.93)(1.267,-0.477){7}{\rule{1.060pt}{0.115pt}}
\multiput(1166.00,647.17)(9.800,-5.000){2}{\rule{0.530pt}{0.400pt}}
\multiput(1178.00,641.94)(1.651,-0.468){5}{\rule{1.300pt}{0.113pt}}
\multiput(1178.00,642.17)(9.302,-4.000){2}{\rule{0.650pt}{0.400pt}}
\multiput(1190.00,637.93)(1.378,-0.477){7}{\rule{1.140pt}{0.115pt}}
\multiput(1190.00,638.17)(10.634,-5.000){2}{\rule{0.570pt}{0.400pt}}
\multiput(1203.00,632.93)(1.267,-0.477){7}{\rule{1.060pt}{0.115pt}}
\multiput(1203.00,633.17)(9.800,-5.000){2}{\rule{0.530pt}{0.400pt}}
\multiput(1215.00,627.94)(1.651,-0.468){5}{\rule{1.300pt}{0.113pt}}
\multiput(1215.00,628.17)(9.302,-4.000){2}{\rule{0.650pt}{0.400pt}}
\multiput(1227.00,623.93)(1.267,-0.477){7}{\rule{1.060pt}{0.115pt}}
\multiput(1227.00,624.17)(9.800,-5.000){2}{\rule{0.530pt}{0.400pt}}
\multiput(1239.00,618.93)(1.378,-0.477){7}{\rule{1.140pt}{0.115pt}}
\multiput(1239.00,619.17)(10.634,-5.000){2}{\rule{0.570pt}{0.400pt}}
\multiput(1252.00,613.94)(1.651,-0.468){5}{\rule{1.300pt}{0.113pt}}
\multiput(1252.00,614.17)(9.302,-4.000){2}{\rule{0.650pt}{0.400pt}}
\multiput(1264.00,609.94)(1.651,-0.468){5}{\rule{1.300pt}{0.113pt}}
\multiput(1264.00,610.17)(9.302,-4.000){2}{\rule{0.650pt}{0.400pt}}
\multiput(1276.00,605.94)(1.797,-0.468){5}{\rule{1.400pt}{0.113pt}}
\multiput(1276.00,606.17)(10.094,-4.000){2}{\rule{0.700pt}{0.400pt}}
\multiput(1289.00,601.94)(1.651,-0.468){5}{\rule{1.300pt}{0.113pt}}
\multiput(1289.00,602.17)(9.302,-4.000){2}{\rule{0.650pt}{0.400pt}}
\multiput(1301.00,597.94)(1.651,-0.468){5}{\rule{1.300pt}{0.113pt}}
\multiput(1301.00,598.17)(9.302,-4.000){2}{\rule{0.650pt}{0.400pt}}
\multiput(1313.00,593.94)(1.651,-0.468){5}{\rule{1.300pt}{0.113pt}}
\multiput(1313.00,594.17)(9.302,-4.000){2}{\rule{0.650pt}{0.400pt}}
\multiput(1325.00,589.95)(2.695,-0.447){3}{\rule{1.833pt}{0.108pt}}
\multiput(1325.00,590.17)(9.195,-3.000){2}{\rule{0.917pt}{0.400pt}}
\multiput(1338.00,586.95)(2.472,-0.447){3}{\rule{1.700pt}{0.108pt}}
\multiput(1338.00,587.17)(8.472,-3.000){2}{\rule{0.850pt}{0.400pt}}
\multiput(1350.00,583.95)(2.472,-0.447){3}{\rule{1.700pt}{0.108pt}}
\multiput(1350.00,584.17)(8.472,-3.000){2}{\rule{0.850pt}{0.400pt}}
\multiput(1362.00,580.95)(2.695,-0.447){3}{\rule{1.833pt}{0.108pt}}
\multiput(1362.00,581.17)(9.195,-3.000){2}{\rule{0.917pt}{0.400pt}}
\multiput(1375.00,577.95)(2.472,-0.447){3}{\rule{1.700pt}{0.108pt}}
\multiput(1375.00,578.17)(8.472,-3.000){2}{\rule{0.850pt}{0.400pt}}
\put(1387,574.17){\rule{2.500pt}{0.400pt}}
\multiput(1387.00,575.17)(6.811,-2.000){2}{\rule{1.250pt}{0.400pt}}
\multiput(1399.00,572.95)(2.472,-0.447){3}{\rule{1.700pt}{0.108pt}}
\multiput(1399.00,573.17)(8.472,-3.000){2}{\rule{0.850pt}{0.400pt}}
\put(1411,569.17){\rule{2.700pt}{0.400pt}}
\multiput(1411.00,570.17)(7.396,-2.000){2}{\rule{1.350pt}{0.400pt}}
\put(1424,567.17){\rule{2.500pt}{0.400pt}}
\multiput(1424.00,568.17)(6.811,-2.000){2}{\rule{1.250pt}{0.400pt}}
\put(1080.0,664.0){\rule[-0.200pt]{2.891pt}{0.400pt}}
\put(220,119){\usebox{\plotpoint}}
\put(220,118.67){\rule{2.891pt}{0.400pt}}
\multiput(220.00,118.17)(6.000,1.000){2}{\rule{1.445pt}{0.400pt}}
\put(245,119.67){\rule{2.891pt}{0.400pt}}
\multiput(245.00,119.17)(6.000,1.000){2}{\rule{1.445pt}{0.400pt}}
\put(257,120.67){\rule{2.891pt}{0.400pt}}
\multiput(257.00,120.17)(6.000,1.000){2}{\rule{1.445pt}{0.400pt}}
\put(232.0,120.0){\rule[-0.200pt]{3.132pt}{0.400pt}}
\put(281,121.67){\rule{3.132pt}{0.400pt}}
\multiput(281.00,121.17)(6.500,1.000){2}{\rule{1.566pt}{0.400pt}}
\put(294,122.67){\rule{2.891pt}{0.400pt}}
\multiput(294.00,122.17)(6.000,1.000){2}{\rule{1.445pt}{0.400pt}}
\put(306,123.67){\rule{2.891pt}{0.400pt}}
\multiput(306.00,123.17)(6.000,1.000){2}{\rule{1.445pt}{0.400pt}}
\put(318,124.67){\rule{3.132pt}{0.400pt}}
\multiput(318.00,124.17)(6.500,1.000){2}{\rule{1.566pt}{0.400pt}}
\put(331,125.67){\rule{2.891pt}{0.400pt}}
\multiput(331.00,125.17)(6.000,1.000){2}{\rule{1.445pt}{0.400pt}}
\put(343,126.67){\rule{2.891pt}{0.400pt}}
\multiput(343.00,126.17)(6.000,1.000){2}{\rule{1.445pt}{0.400pt}}
\put(355,127.67){\rule{2.891pt}{0.400pt}}
\multiput(355.00,127.17)(6.000,1.000){2}{\rule{1.445pt}{0.400pt}}
\put(367,128.67){\rule{3.132pt}{0.400pt}}
\multiput(367.00,128.17)(6.500,1.000){2}{\rule{1.566pt}{0.400pt}}
\put(380,130.17){\rule{2.500pt}{0.400pt}}
\multiput(380.00,129.17)(6.811,2.000){2}{\rule{1.250pt}{0.400pt}}
\put(392,131.67){\rule{2.891pt}{0.400pt}}
\multiput(392.00,131.17)(6.000,1.000){2}{\rule{1.445pt}{0.400pt}}
\put(404,133.17){\rule{2.700pt}{0.400pt}}
\multiput(404.00,132.17)(7.396,2.000){2}{\rule{1.350pt}{0.400pt}}
\put(417,134.67){\rule{2.891pt}{0.400pt}}
\multiput(417.00,134.17)(6.000,1.000){2}{\rule{1.445pt}{0.400pt}}
\put(429,136.17){\rule{2.500pt}{0.400pt}}
\multiput(429.00,135.17)(6.811,2.000){2}{\rule{1.250pt}{0.400pt}}
\put(441,138.17){\rule{2.500pt}{0.400pt}}
\multiput(441.00,137.17)(6.811,2.000){2}{\rule{1.250pt}{0.400pt}}
\put(453,140.17){\rule{2.700pt}{0.400pt}}
\multiput(453.00,139.17)(7.396,2.000){2}{\rule{1.350pt}{0.400pt}}
\multiput(466.00,142.61)(2.472,0.447){3}{\rule{1.700pt}{0.108pt}}
\multiput(466.00,141.17)(8.472,3.000){2}{\rule{0.850pt}{0.400pt}}
\put(478,145.17){\rule{2.500pt}{0.400pt}}
\multiput(478.00,144.17)(6.811,2.000){2}{\rule{1.250pt}{0.400pt}}
\multiput(490.00,147.61)(2.695,0.447){3}{\rule{1.833pt}{0.108pt}}
\multiput(490.00,146.17)(9.195,3.000){2}{\rule{0.917pt}{0.400pt}}
\multiput(503.00,150.61)(2.472,0.447){3}{\rule{1.700pt}{0.108pt}}
\multiput(503.00,149.17)(8.472,3.000){2}{\rule{0.850pt}{0.400pt}}
\multiput(515.00,153.61)(2.472,0.447){3}{\rule{1.700pt}{0.108pt}}
\multiput(515.00,152.17)(8.472,3.000){2}{\rule{0.850pt}{0.400pt}}
\multiput(527.00,156.61)(2.472,0.447){3}{\rule{1.700pt}{0.108pt}}
\multiput(527.00,155.17)(8.472,3.000){2}{\rule{0.850pt}{0.400pt}}
\multiput(539.00,159.60)(1.797,0.468){5}{\rule{1.400pt}{0.113pt}}
\multiput(539.00,158.17)(10.094,4.000){2}{\rule{0.700pt}{0.400pt}}
\multiput(552.00,163.60)(1.651,0.468){5}{\rule{1.300pt}{0.113pt}}
\multiput(552.00,162.17)(9.302,4.000){2}{\rule{0.650pt}{0.400pt}}
\multiput(564.00,167.60)(1.651,0.468){5}{\rule{1.300pt}{0.113pt}}
\multiput(564.00,166.17)(9.302,4.000){2}{\rule{0.650pt}{0.400pt}}
\multiput(576.00,171.60)(1.651,0.468){5}{\rule{1.300pt}{0.113pt}}
\multiput(576.00,170.17)(9.302,4.000){2}{\rule{0.650pt}{0.400pt}}
\multiput(588.00,175.59)(1.378,0.477){7}{\rule{1.140pt}{0.115pt}}
\multiput(588.00,174.17)(10.634,5.000){2}{\rule{0.570pt}{0.400pt}}
\multiput(601.00,180.59)(1.267,0.477){7}{\rule{1.060pt}{0.115pt}}
\multiput(601.00,179.17)(9.800,5.000){2}{\rule{0.530pt}{0.400pt}}
\multiput(613.00,185.59)(1.267,0.477){7}{\rule{1.060pt}{0.115pt}}
\multiput(613.00,184.17)(9.800,5.000){2}{\rule{0.530pt}{0.400pt}}
\multiput(625.00,190.59)(1.123,0.482){9}{\rule{0.967pt}{0.116pt}}
\multiput(625.00,189.17)(10.994,6.000){2}{\rule{0.483pt}{0.400pt}}
\multiput(638.00,196.59)(1.033,0.482){9}{\rule{0.900pt}{0.116pt}}
\multiput(638.00,195.17)(10.132,6.000){2}{\rule{0.450pt}{0.400pt}}
\multiput(650.00,202.59)(1.033,0.482){9}{\rule{0.900pt}{0.116pt}}
\multiput(650.00,201.17)(10.132,6.000){2}{\rule{0.450pt}{0.400pt}}
\multiput(662.00,208.59)(0.874,0.485){11}{\rule{0.786pt}{0.117pt}}
\multiput(662.00,207.17)(10.369,7.000){2}{\rule{0.393pt}{0.400pt}}
\multiput(674.00,215.59)(0.950,0.485){11}{\rule{0.843pt}{0.117pt}}
\multiput(674.00,214.17)(11.251,7.000){2}{\rule{0.421pt}{0.400pt}}
\multiput(687.00,222.59)(0.874,0.485){11}{\rule{0.786pt}{0.117pt}}
\multiput(687.00,221.17)(10.369,7.000){2}{\rule{0.393pt}{0.400pt}}
\multiput(699.00,229.59)(0.758,0.488){13}{\rule{0.700pt}{0.117pt}}
\multiput(699.00,228.17)(10.547,8.000){2}{\rule{0.350pt}{0.400pt}}
\multiput(711.00,237.59)(0.824,0.488){13}{\rule{0.750pt}{0.117pt}}
\multiput(711.00,236.17)(11.443,8.000){2}{\rule{0.375pt}{0.400pt}}
\multiput(724.00,245.59)(0.758,0.488){13}{\rule{0.700pt}{0.117pt}}
\multiput(724.00,244.17)(10.547,8.000){2}{\rule{0.350pt}{0.400pt}}
\multiput(736.00,253.59)(0.669,0.489){15}{\rule{0.633pt}{0.118pt}}
\multiput(736.00,252.17)(10.685,9.000){2}{\rule{0.317pt}{0.400pt}}
\multiput(748.00,262.59)(0.669,0.489){15}{\rule{0.633pt}{0.118pt}}
\multiput(748.00,261.17)(10.685,9.000){2}{\rule{0.317pt}{0.400pt}}
\multiput(760.00,271.58)(0.652,0.491){17}{\rule{0.620pt}{0.118pt}}
\multiput(760.00,270.17)(11.713,10.000){2}{\rule{0.310pt}{0.400pt}}
\multiput(773.00,281.59)(0.669,0.489){15}{\rule{0.633pt}{0.118pt}}
\multiput(773.00,280.17)(10.685,9.000){2}{\rule{0.317pt}{0.400pt}}
\multiput(785.00,290.58)(0.600,0.491){17}{\rule{0.580pt}{0.118pt}}
\multiput(785.00,289.17)(10.796,10.000){2}{\rule{0.290pt}{0.400pt}}
\multiput(797.00,300.58)(0.652,0.491){17}{\rule{0.620pt}{0.118pt}}
\multiput(797.00,299.17)(11.713,10.000){2}{\rule{0.310pt}{0.400pt}}
\multiput(810.00,310.58)(0.600,0.491){17}{\rule{0.580pt}{0.118pt}}
\multiput(810.00,309.17)(10.796,10.000){2}{\rule{0.290pt}{0.400pt}}
\multiput(822.00,320.59)(0.669,0.489){15}{\rule{0.633pt}{0.118pt}}
\multiput(822.00,319.17)(10.685,9.000){2}{\rule{0.317pt}{0.400pt}}
\multiput(834.00,329.59)(0.669,0.489){15}{\rule{0.633pt}{0.118pt}}
\multiput(834.00,328.17)(10.685,9.000){2}{\rule{0.317pt}{0.400pt}}
\multiput(846.00,338.59)(0.728,0.489){15}{\rule{0.678pt}{0.118pt}}
\multiput(846.00,337.17)(11.593,9.000){2}{\rule{0.339pt}{0.400pt}}
\multiput(859.00,347.59)(0.669,0.489){15}{\rule{0.633pt}{0.118pt}}
\multiput(859.00,346.17)(10.685,9.000){2}{\rule{0.317pt}{0.400pt}}
\multiput(871.00,356.59)(0.874,0.485){11}{\rule{0.786pt}{0.117pt}}
\multiput(871.00,355.17)(10.369,7.000){2}{\rule{0.393pt}{0.400pt}}
\multiput(883.00,363.59)(0.950,0.485){11}{\rule{0.843pt}{0.117pt}}
\multiput(883.00,362.17)(11.251,7.000){2}{\rule{0.421pt}{0.400pt}}
\multiput(896.00,370.59)(1.033,0.482){9}{\rule{0.900pt}{0.116pt}}
\multiput(896.00,369.17)(10.132,6.000){2}{\rule{0.450pt}{0.400pt}}
\multiput(908.00,376.60)(1.651,0.468){5}{\rule{1.300pt}{0.113pt}}
\multiput(908.00,375.17)(9.302,4.000){2}{\rule{0.650pt}{0.400pt}}
\multiput(920.00,380.61)(2.472,0.447){3}{\rule{1.700pt}{0.108pt}}
\multiput(920.00,379.17)(8.472,3.000){2}{\rule{0.850pt}{0.400pt}}
\put(932,382.67){\rule{3.132pt}{0.400pt}}
\multiput(932.00,382.17)(6.500,1.000){2}{\rule{1.566pt}{0.400pt}}
\put(269.0,122.0){\rule[-0.200pt]{2.891pt}{0.400pt}}
\put(957,382.17){\rule{2.500pt}{0.400pt}}
\multiput(957.00,383.17)(6.811,-2.000){2}{\rule{1.250pt}{0.400pt}}
\multiput(969.00,380.93)(1.378,-0.477){7}{\rule{1.140pt}{0.115pt}}
\multiput(969.00,381.17)(10.634,-5.000){2}{\rule{0.570pt}{0.400pt}}
\multiput(982.00,375.93)(1.033,-0.482){9}{\rule{0.900pt}{0.116pt}}
\multiput(982.00,376.17)(10.132,-6.000){2}{\rule{0.450pt}{0.400pt}}
\multiput(994.00,369.93)(0.758,-0.488){13}{\rule{0.700pt}{0.117pt}}
\multiput(994.00,370.17)(10.547,-8.000){2}{\rule{0.350pt}{0.400pt}}
\multiput(1006.00,361.92)(0.600,-0.491){17}{\rule{0.580pt}{0.118pt}}
\multiput(1006.00,362.17)(10.796,-10.000){2}{\rule{0.290pt}{0.400pt}}
\multiput(1018.00,351.92)(0.539,-0.492){21}{\rule{0.533pt}{0.119pt}}
\multiput(1018.00,352.17)(11.893,-12.000){2}{\rule{0.267pt}{0.400pt}}
\multiput(1031.58,338.79)(0.492,-0.539){21}{\rule{0.119pt}{0.533pt}}
\multiput(1030.17,339.89)(12.000,-11.893){2}{\rule{0.400pt}{0.267pt}}
\multiput(1043.58,325.51)(0.492,-0.625){21}{\rule{0.119pt}{0.600pt}}
\multiput(1042.17,326.75)(12.000,-13.755){2}{\rule{0.400pt}{0.300pt}}
\multiput(1055.58,310.54)(0.493,-0.616){23}{\rule{0.119pt}{0.592pt}}
\multiput(1054.17,311.77)(13.000,-14.771){2}{\rule{0.400pt}{0.296pt}}
\multiput(1068.58,294.23)(0.492,-0.712){21}{\rule{0.119pt}{0.667pt}}
\multiput(1067.17,295.62)(12.000,-15.616){2}{\rule{0.400pt}{0.333pt}}
\multiput(1080.58,277.09)(0.492,-0.755){21}{\rule{0.119pt}{0.700pt}}
\multiput(1079.17,278.55)(12.000,-16.547){2}{\rule{0.400pt}{0.350pt}}
\multiput(1092.58,258.96)(0.492,-0.798){21}{\rule{0.119pt}{0.733pt}}
\multiput(1091.17,260.48)(12.000,-17.478){2}{\rule{0.400pt}{0.367pt}}
\multiput(1104.58,240.16)(0.493,-0.734){23}{\rule{0.119pt}{0.685pt}}
\multiput(1103.17,241.58)(13.000,-17.579){2}{\rule{0.400pt}{0.342pt}}
\multiput(1117.58,220.96)(0.492,-0.798){21}{\rule{0.119pt}{0.733pt}}
\multiput(1116.17,222.48)(12.000,-17.478){2}{\rule{0.400pt}{0.367pt}}
\multiput(1129.58,201.82)(0.492,-0.841){21}{\rule{0.119pt}{0.767pt}}
\multiput(1128.17,203.41)(12.000,-18.409){2}{\rule{0.400pt}{0.383pt}}
\multiput(1141.58,181.96)(0.492,-0.798){21}{\rule{0.119pt}{0.733pt}}
\multiput(1140.17,183.48)(12.000,-17.478){2}{\rule{0.400pt}{0.367pt}}
\multiput(1153.58,163.03)(0.493,-0.774){23}{\rule{0.119pt}{0.715pt}}
\multiput(1152.17,164.52)(13.000,-18.515){2}{\rule{0.400pt}{0.358pt}}
\multiput(1166.58,142.96)(0.492,-0.798){21}{\rule{0.119pt}{0.733pt}}
\multiput(1165.17,144.48)(12.000,-17.478){2}{\rule{0.400pt}{0.367pt}}
\multiput(1178.59,124.00)(0.489,-0.786){15}{\rule{0.118pt}{0.722pt}}
\multiput(1177.17,125.50)(9.000,-12.501){2}{\rule{0.400pt}{0.361pt}}
\put(945.0,384.0){\rule[-0.200pt]{2.891pt}{0.400pt}}
\sbox{\plotpoint}{\rule[-0.400pt]{0.800pt}{0.800pt}}%
\put(757,300){\raisebox{-.8pt}{\makebox(0,0){$\Diamond$}}}
\put(864,446){\raisebox{-.8pt}{\makebox(0,0){$\Diamond$}}}
\put(972,617){\raisebox{-.8pt}{\makebox(0,0){$\Diamond$}}}
\put(1079,695){\raisebox{-.8pt}{\makebox(0,0){$\Diamond$}}}
\put(1187,684){\raisebox{-.8pt}{\makebox(0,0){$\Diamond$}}}
\put(1294,656){\raisebox{-.8pt}{\makebox(0,0){$\Diamond$}}}
\put(1401,639){\raisebox{-.8pt}{\makebox(0,0){$\Diamond$}}}
\put(366,801){\raisebox{-.8pt}{\makebox(0,0){$\Diamond$}}}
\sbox{\plotpoint}{\rule[-0.500pt]{1.000pt}{1.000pt}}%
\put(435,114){\makebox(0,0){$+$}}
\put(542,121){\makebox(0,0){$+$}}
\put(650,144){\makebox(0,0){$+$}}
\put(757,221){\makebox(0,0){$+$}}
\put(864,401){\makebox(0,0){$+$}}
\put(972,603){\makebox(0,0){$+$}}
\put(1079,688){\makebox(0,0){$+$}}
\put(1187,680){\makebox(0,0){$+$}}
\put(1294,655){\makebox(0,0){$+$}}
\put(1401,638){\makebox(0,0){$+$}}
\put(366,750){\makebox(0,0){$+$}}
\end{picture}
\caption{The mass of the Schwinger boson:
Our first and second order results, along
with lattice and DLCQ calculations.}
\label{comparison}
\end{figure}
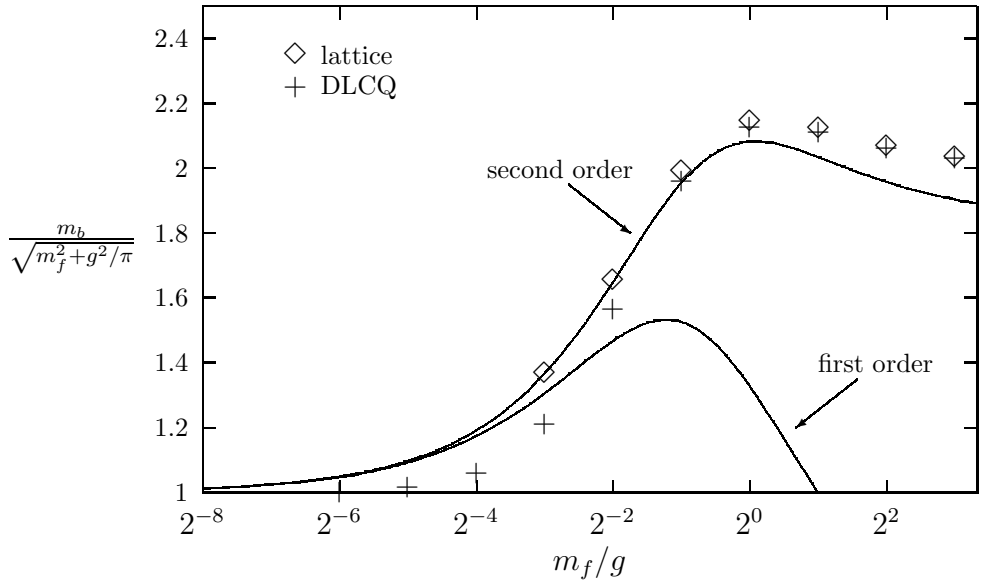

A plot of this second order result, our first order
result, a lattice result$^5$, and a DLCQ result$^6$ is
shown in figure \ref{comparison}.
There are several things to notice about the plot.
\begin{itemize}
\item{Our results through second order match the lattice results
for an intermediate range of
$m_f/m_b^{(0)}$, and we extend the lattice results to
stronger coupling (lighter mass).}
\item{DLCQ misses the linear contribution to the mass:  the leading
term in a DLCQ calculation is $\propto m_f^2$.}
\end{itemize}

{\bf 3. Structure Functions}

Our structure functions will depend on a parameter $K$:
\begin{equation}
K = \frac{p_{-}}{m_b\left(\epsilon/L\right )
^\frac{1}{2}}
\end{equation}
This parameter fixes our frame, and we are interested in results
as $K \to \infty$.
There are two ways we could obtain this limit:
\begin{itemize}
\item{$p_{-} \to \infty$.  This is reminiscent of an
infinite momentum frame scheme.}
\item{$\epsilon/L \to 0$.  This reduces our coordinates
to the usual light front coordinates.}
\end{itemize}
In this way, the parameter $K$ forms an useful bridge
between the infinite momentum frame method and light
front coordinates.

At the moment, the `full' structure function result
(with arbitary numbers of intermediate bosons) seems to be
numerically challenging.
Calculations to date have concentrated on the lowest 3-boson
contribution to the structure functions.
Preliminary results show a significantly asymmetric distribution
around $x=0.5$, whereas DLCQ and wave equation approaches
(through $qq\bar{q}\bar{q}$ configurations) show symmetric
distrbutions.
Calculations are in progress to obtain the entire summed series.

\vskip.3in
{\bf Acknowledgements}
\vskip.1in
This work was supported in part by the U.S. Department of Energy
under Grant No. DEFG02-87ER40371, Division of High Energy and Nuclear
Physics.  JPV acknowledges support from the Alexander von Humboldt
Foundation.  We acknowledge valuable discussions with M.~Burkardt,
P.~Griffin, and H.C.~Pauli.
\vskip.3in
{\bf References}
\vskip.1in

1. E.V.~Prokhvatilov and V.A.~Franke, Sov.\ J.\ Nucl.\ Phys.\
   {\bf 49}, 688 (1989).

2. F.~Lenz, M.~Thies, S.~Levit and K.~Yazaki, Ann.\ Phys.\
   {\bf 208}, 1 (1991).

3. M.~Burkardt, Phys.\ Rev.\ {\bf D47}, 4628 (1993);
   Habilitationsschrift, Erlangen, (1994).

4. P.A.~Griffin, Phys.\ Rev.\ {\bf D46}, 3538 (1992).

5. D.P.~Crewther and C.J.~Hamer, Nucl.\ Phys.\ {\bf B170}
   , 353 (1980).

6. S.~Elser, Diploma Thesis, Heidelberg (1994)
\end{document}